Title:    CONSCIOUSNESS: A DIRECT LINK WITH LIFE'S ORIGIN?

Authors:   A. N. Mitra[1]    and    G. Mitra-Delmotte[2], Ph.D.

[1]Emeritus, Department of Physics, Delhi University, Delhi-110007; 244 Tagore Park, Delhi-110009; e.mail: ganmitra@nde.vsnl.net.in

[2]39 Cite de l'Ocean, Montgaillard, 97400 St.Denis, Reunion Island ;
e.mail: gargijj@orange.fr

**Abstract**: Inspired by the Penrose-Hameroff thesis, we are intuitively led to examine an intriguing correspondence of 'induction' (by fields), with the complex phenomenon of (metabolism-sustained) consciousness: Did sequences of associated induction patterns in field-susceptible biomatter have simpler beginnings?

**Keywords**: mind-matter; field-driven assembly; induction; environment; solitons; coherence; magnetism.

1. Introduction

Consciousness is a many-splendoured thing, whose anatomy has been under scrutiny almost since the birth of civilization. We have come a long way from the times when this term was associated with religion and spiritualism, to the present era when serious efforts are directed towards understanding it in the language of Science (Penrose 1995; Hameroff 2003). During this saga, physical science has progressed all the way from Newtonian mechanics (when Cartesian Partition ruled against such efforts) to the birth of relativity and quantum mechanics when sheer compulsions of logical self-consistency demanded that Cartesian Partition was no longer tenable and that  mind and matter could no longer be divorced from each other. This was despite Bohr's Copenhagen Interpretation which had effectively decreed against difficult logical questions being asked about quantum mechanics. But Einstein could not accept this dictum and produced his EPR paper (Einstein et al 1935) ostensibly designed to demolish the tenets of quantum mechanics, but serendipitously treaded on a most fertile land as a logical consequence of the new paradigm, namely *quantum entanglement* and *non-locality*. This was directly at variance with the concept of *local realism*, the bedrock of the Copenhagen Interpretation. Since both could not be right at the same time, it took another half a century to decide on the issue: Alan Aspect, through his famous experiment (Aspect et al 1982), gave the verdict in favour of EPR's entanglement and non-locality and in so doing, ruled out Bohr's local causality. In the meantime Copenhagen had got rather outdated due to the emergence of *decoherence,* thanks to a two-decade-old development bearing on the very foundations of quantum mechanics (following the Aspect discovery) which gave an increasingly active role to the *environment* (see below for its fuller ramifications).

This episode offers a possible setting for bringing mind and matter on a common platform, since a direct touch with reality of these bizarre quantum concepts have willy-nilly got these two entities "mutually entangled"! One may wonder how this 'Frankenstein' (read science with all its tools) which is the product of the human mind in the first place, has come to challenge its own Creator, and probe its `anatomy'.

This essay is an attempt to string together some scientific advances designed to reduce the complex phenomena of consciousness (Sect.2-5), and to map (Sect.6-7) the resulting scenario to simpler ingredients that may have been available in the Hadean.

## 2. Bohm's Thesis: Integral Duality of Mind and Matter

For historical reasons, we start with a semi-intuitive model due to David Bohm (1990) who was led, by the conflict of quantum theory (discreteness) with GTR (continuous matter), to propose the existence of an undivided wholeness present in an *implicate* order which applies to both matter and mind, so that it can in principle access the relationship between these two different things. In this picture, matter and mind are seen as relative projections into an explicate order from the reality of the implicate order, with apparently no connection between them. Only at the deeper, fundamental level of the universe, does there exist an unbroken wholeness in which mind (consciousness?) merges with matter-- something akin to a holographic image of the brain (Pribram 1975). Bohm also illustrates the idea of `meaning' through the example of listening to music as a sequence of overlapping moments each with a short but finite interval of time. To produce the notes, one moment 'induces' the next, such that the content that was implicate in the immediate past, becomes explicate in the next interval (the immediate present). The sense of movement in music is thus the result of a sequence of overlapping transitions, thus producing consciousness from an implicate order. Consciousness is thus seen to be intimately related to the concept of `time' –not merely a 'duration', in the sense of classical mechanics, but an active ingredient bearing on consciousness that reveals a world of continuous and unfolding events, *a la* Bergson (2001[1889]). Bohm (1990) further suggests that the tiny electron is an inseparable union of particle *and* (*not or*) field wherein the latter (like consciousness) organizes the movement of the former (like body). This is in line with Bohm's (1952) earlier thesis on quantum mechanics with

"hidden variables" wherein guiding waves determine the motion of the associated particles.

**3. Quantum Coherence in Biology**

Next, quantum coherence, which is naturally associated with quantum mechanics, is considered as a key ingredient in a more recent approach to consciousness studies, questioning the validity of purely algorithmic prescriptions for addressing phenomena like human insight. To deal with this issue of "non-computability", Penrose (1995) suggests a new role for the *environment*, viz., as an "external guide" influencing decisions in an algorithmic system, and for this a necessary condition is its quantum coherence. For a more concrete representation of such a picture in a biosystem, let us look at Frohlich's (1968) coherent excitations envisaged for cell membranes. Now, a stationary state is reached if the energy fed into an assembled material with polar vibrational modes, is sufficiently larger than that lost to bath degrees of freedom. Then, in the words of Frohlich (1968), "The long-range Coulomb interaction then causes this energy to be shared with other dipoles. …the dipoles will tend to oscillate coherently provided the energy supply is sufficiently large compared with the energy loss. Non-linear effects are likely to reduce this loss with increasing excitation and effectively transfer the system into a metastable state in which the energy supplied locally to dipolar constituents is channelled into a single longitudinal mode which exhibits long-range phase correlations". Of course, the energy of the metabolic drive must be large enough, and the dielectric properties of the materials concerned need to have a matching capacity for maintaining (and withstanding) extremely high electric fields (Frohlich 1975). This is similar to Bose-Einstein condensation, in which a large number of particles participate in a *single quantum state*, i.e., behaving as one with a wave function applicable for a single particle, albeit scaled up appropriately. Despite their much higher than 'absolute zero' temperatures, it seems that coherence conditions are not only met in bio-systems, but that there is some direct experimental evidence too (Grundler and Keilmann 1983) of $10^{11}$ Hz oscillations predicted by Frohlich (1968). If this seems surprising vis-a-vis physical systems, note that it is *because of* (not despite!) the warm and soft sol-gel state that efficient nano-machines actually harness thermal fluctuations (Bustamante et al 2005, see below) in biology, where energy transformations occur under isothermal conditions.

## 4. Penrose-Hameroff Model: New Role for the Environment

In looking for appropriate biomaterials that on supplying with energy led to Frohlich-like excitations in the sub-nanosecond range, the Penrose-Hameroff model zeroes in on microtubules having the required ingredients of voltage effects and a geometry favoring "coupling among subunits" (Hameroff 2003) for acting as reservoirs of highly ordered energy. The early-evolved cytoskeleton forms the basic 'building block' in their new fractal perspective of the nervous system, which argues for a change in paradigm so that the sophisticated actions of animals down to single celled ones (all affected by general anaesthetics at about the same concentration) can be explained using only one basic control system (Penrose 1995). Now *functional* protein conformations appear to be correlated to such collective "metastable states" mediated via the surroundings: "action of electric fields, binding of ligands or neurotransmitters, or effects of neighbor proteins" (Hameroff 2003). Thus consciousness can be partially inhibited using anaesthetics. (By reversibly binding to hydrophobic pockets within key neural proteins via weak forces, these can alter the environmental medium and thereby the electron mobility, in turn non-linearly coupled to mechanical movements).

A complementary approach to such long range cooperativity is due to Davydov (see Lomdahl et al 1984) who considers wave-like propagations—solitons-- for the spatial transfer of vibrational energy in ordered form, which also can be derived from the same type of non-linear effects leading to the coherent Frohlich ordered state. Tuszynski et al (1984) observe that while the latter lays its emphasis on time-independent dynamical ordering aspects, the former offers a plausible mechanism for not only localization but also transporting order through the system (time dependent aspects). (The unusual resilience of a soliton-- a quantum of energy that propagates as a traveling wave in nonlinear systems-- stems from two opposing tendencies as a result of which a dynamically stable entity emerges). Indeed, as the substrate for energy transfer in the cytoskeleton, just like electrons in computers, Hameroff (2003) considers Davydov's solitons. "Objective reduction" (a self-collapse of the observer's wave function), then occurs in this algorithmic coherent system (see above), one in which the external (gravitational) field plays a key role. Here, the Penrose-Hameroff thesis makes a major departure from the conventional view, in that the (non-computable) field shows a new and active role for the *environment*, viz., as a concerned 'teacher' with a deep involvement in the system's decision-making- not merely a neutral examiner, assessing system variants (Penrose 1995).

Further, another study (Davia 2006) suggests the relation between the organism and the environment as one of mutual 'friendliness', in contrast to the reigning Darwinian perspective where, apart from being a source/sink, the environment presents itself to living systems as a sort of (potentially destructive) obstacle course to be negotiated; and the organism appears as a machine *within* an environment, with no causal relationship between the two. Briefly, Davia seeks to demonstrate that the question of how life maintains its organization through time is central to an understanding of the brain. To that end, he postulates life to be a scale- free (fractal) process of catalysis (which involves the fusion of energy and structure in the form of solitons). Then, rather than a hostile 'obstacle course', the environment becomes a willing partner in a set of transitions mediated by the living process via `catalysis'.

**5. Non-computable Bio-issues: External Control?**

According to *Goedel's incompleteness theorem*, with any set of axioms, it is possible to produce a statement that is obviously true, but cannot be proved by means of the axioms themselves. Penrose (1995) took advantage of the Goedel theorem to claim that the functioning of the human brain also includes *non-algorithmic* processes, i.e. a system can be deterministic without being algorithmic. For this an excellent candidate is again quantum mechanics in its full glory of *quantum coherence*, in common with Bohm's (1990) semi-intuitive thesis. Now, mathematical induction is a well known concept which is akin to Goedel's theorem. Inspired by Penrose, we wish to extend this terminology to a *physical* level via the well known phenomena of (electric, magnetic) field-induced effects, which although conceived classically has a good promise of quantum extension. And with due respect to gravitation, the effects of other fields at different levels, from classical to quantum, should not be neglected in view of the properties of biomatter (Cope 1975).

We again return to the theme of non-computability for thought processes (Penrose 1995), looking to the environment as exerting external control. Now in fact, across biology we encounter instances where bio-solutions can include choices *outside* the space of existing possibilities. For instance, consider Bio-evolution (in particular the algorithmic complexity of sequences pointed out by Abel (2009)), and note that a similar Darwinian selection at the time scale of days--affinity maturation in B-cells-- can be found in higher vertebrates. So it is natural to ask how life itself must have emerged, (very likely from a set of not-as-yet-living systems), thus taking the problem to the door of life's origin!

**6. The Environment as Guide; a Scaffold for Life (?)**

Now in addition to the vital role for solitons in today's biology, they have strong implications for life's emergence owing to their fundamental association with both energy and information (Taranenko et al 2005). That is the boundary conditions offered by repeating structures could have been the answer to how energy and structure in biology got synonymous (Davia 2006), so that the patterns sustaining these quasiparticles could have been retained while gradually replacing the materials embodying them, *en route* to present day versions of metabolism and replication (c.f. Cairns Smith 1982; see below). In this context, Davydov's (1991) proposal for 'electrosolitons' (a plausible mechanism for electron transfers across distances with minimum energy losses, traditionally approached using tunneling effects) seems to be highly relevant for the hydrogenation of $CO_2$--seen as the basis for life's emergence (Nitschke and Russell 2009). Indeed, the quantum metabolism model of Demetrius (2008) approaches the issue of energy harnessing in the ATP-membrane proton pump-the most primitive of energy transduction mechanisms-using Frohlich's coherent excitations. Hence, revisiting Cairns-Smith's (1982) idea of a mineral scaffold for life 'taken over' by organic matter, in the light of these insights, prompts the question of whether the above non-linear interactions leading to coherent dynamics could have been achieved using simpler/less sophisticated substances that in turn got gradually replaced by the advanced ones of today with greater time-stability. Importantly, the new ingredient would be the environment having a penetrating influence in the coherent scaffold, and taking decisions *a la* Penrose.

A few years ago, we have drawn attention to another ubiquitous ingredient in terrestrial phenomena, viz., magnetism, which appears to exert its influence across kingdoms of life, and has a natural association with quantum *coherence* (see Merali 2007). A soft colloidal scaffold *a la* Russell and coworkers (1989) in terms of a field-induced assembly of magnetic dipoles (Mitra-Delmotte and Mitra 2010b), seems equipped with the potential to address symmetry-broken dynamics for primordial chemical reactions hosted within its 'layers'. A magnetic field can 'order' magnetic nanoparticles; the resulting structural order in natural assemblies could provide the boundary conditions needed for generating soliton-like structures. The synonymy between structure and energy across biology (Davia 2006) makes it compelling to speculate if magnetic solitons could not have been a primitive mechanism (c.f. Cairns-Smith 2008) for energy transport in a natural assembly (Russell et al 1990; Sawlowicz 1983), whose dynamical order was controlled by a field. To that end, it is encouraging to find studies using particles interacting via dipolar interactions (Ishizaka and Nakamura 2000) and indeed worth noting the recent field-modulated dynamics of magnetic nanoparticle ensembles by Casic et al. (2010). That solitons could be linked to transfer of order within field-induced colloidal structures, shows to what extent the analogies of energy landscapes for protein-folding and of disordered (solid) spin systems can be extended, and which thereby reduces the immense gap demarcating living and (considered as) non-living matter. Then it becomes tempting to cite a few other apparently disjointed features which fit into a bigger mosaic. For instance, there are intriguing analogies of conformational fluctuations of 'sophisticated' motor proteins carrying a load, with infinitesimal spin alignment changes of magnetic dipoles, ligand bound to organics, making their way through templates of head-to-tail aligned electric and magnetic dipoles, respectively (Mitra-Delmotte and Mitra 2010a). These can throw light on how thermal fluctuations can be harnessed in a simpler system with such life-like features, and which seems plausible to imagine in a Hadean setting. Like in ATP-driven molecular motors, a gentle flux gradient (in a non-homogeneous rock magnetic field) can offer both detailed-balance breaking non-equilibrium as well as asymmetry to a magnetic dipole. Again, the correspondence of the local lowering of temperature (towards aiding coherence) theorized by Matsuno and Paton (2000) via the slow release mechanism of ATP hydrolysis, to the magnetic scenario comes in the form of an accompanying magnetocaloric effect, which allows interchange between system-entropy and bath temperature. And, not only does the matter-structuring role of a magnetic field gel with the boundary requirements of soliton-like structures, it provides a friendly background for a more dynamic role mediated by the soliton, besides being the *same* ingredient already found to be crucial to the *Frohlich mechanism* (1968).

**7. Any Direct Link to the Origins?**

Indeed, the above *inductive* form of reasoning by analogies, which is complementary to algorithmic *deduction* (c.f. Penrose 1995), matches the traditional 'pattern-recognition' approach to biology. Guided by the above, and the repetitively appearing phenomenon of sequential induction (in association with a self-referential character) across a hierarchy of life-processes, we propose that associated induction patterns could offer a richer 'simulation' of an 'active' experience as compared to a mere collection of data on a screen by a computer programmed to 'passively' mimic the same.

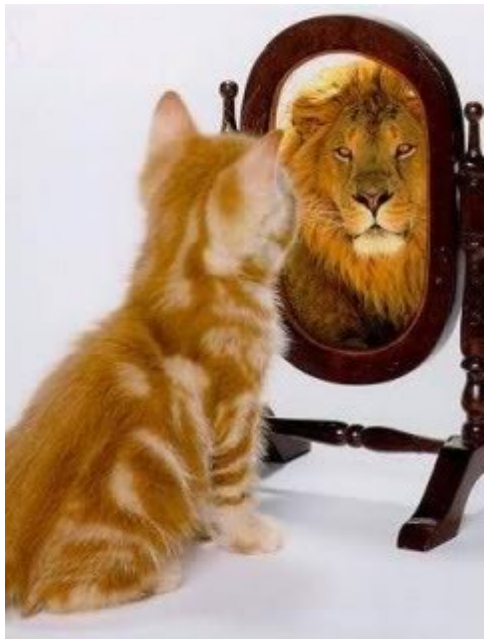

Figure 1:   The subjective experience

The picture above (Figure 1) (taken from the Web) depicts the said subjective experience. Indeed, the communication of the image via propagating patterns *induced* within a biological device could be what makes room for optical illusion effects and subjectivity. Now contrast this typical example of observation/measurement in biology with a detector

(without outside field effects), say a camera, where the corresponding information gets *quenched* on the image plate. So this begs the question whether sequential induction as a measuring mode could throw light on the origins of the complex phenomena of consciousness, in view of the field-susceptible nature of biological matter. Note further that an environment--as a field-- does have the potential for induction, given access to any *active d.o.f.s* in matter. Such a scenario seems to gel with our proposal of a field-induced primitive scaffold for life.

**8. Conclusion: Extra Scientific Dimensions?**

We have chosen only a few samples of consciousness models, at the same time trying to extrapolate the environment-related issues to the emergence of life itself, yet they seem to go hardly beyond scratching only the outer surfaces of the problem so far. The huge gap is perhaps symbolized by a perspective taken from Whitehead if one substitutes "consciousness" for his definition of "religion" (Whitehead 1970):

"Religion is the vision of something which stands beyond, behind, and within the passing flux of immediate things; something which is real, and yet waiting to be realized; something which is a remote possibility, and yet the greatest of present facts; something which gives meaning to all that passes, and yet eludes apprehension; something whose possession is the final good, and yet is beyond all reach; something which is the ultimate ideal, and the hopeless quest."

**Acknowledgements**: This essay is dedicated to the memory of ANM's mother, Rama Rani Mitra, on the occasion of her birth centenary (2011). We thank Prof.M.J.Russell for inspiration and constant help; and Dr.J.J. Delmotte for financial/infrastructural support.